\documentclass[11pt]{article}
\usepackage{times}
\usepackage{geometry}
\usepackage[utf8]{inputenc}
\usepackage{enumitem,amssymb}
\usepackage{ragged2e}
\newlist{thematic}{itemize}{8}
\setlist[thematic]{label=$\square$}
\usepackage{pifont}
\usepackage{fancyhdr}
\usepackage[USenglish]{babel}
\usepackage[utf8]{inputenc}
\usepackage[T1]{fontenc}
\usepackage{epsfig}
\usepackage{wrapfig}
\usepackage{color}

\usepackage[vflt]{floatflt}
\usepackage{longtable}
\usepackage{caption}
\usepackage{jheppub}   
\usepackage[dvipsnames,svgnames,x11names]{xcolor}
\usepackage[all]{hypcap} 
\usepackage{amssymb,amsmath}
\usepackage{amssymb}
\usepackage{amsmath, xspace}
\usepackage{graphics,graphicx} 
\usepackage{rotating}
\usepackage{multirow}
\usepackage{subfigure}
\usepackage{sectsty}
\usepackage[outercaption]{sidecap}
\sidecaptionvpos{figure}{c}

\definecolor{DarkGreen}{rgb}{0.0, 0.3, 0.0}
\definecolor{purple}{rgb}{0.5, 0.0, 0.5}
\definecolor{red}{rgb}{1, 0.0, 0.0}
\definecolor{green}{rgb}{0, 1.0, 0.0}

\def\3he{$^3{\rm He}$}

\def\lsim{\mathrel{\lower2.5pt\vbox{\lineskip=0pt\baselineskip=0pt
           \hbox{$<$}\hbox{$\sim$}}}}

\def\gsim{\mathrel{\lower2.5pt\vbox{\lineskip=0pt\baselineskip=0pt
           \hbox{$>$}\hbox{$\sim$}}}}

\begin{document}
\huge
\begin{center}
AtLAST - Determination of Halo Mass Density Profiles \\ at kpc Scales through Magnification Bias\\
{\normalsize This white paper was submitted to ESO Expanding Horizons in support of AtLAST}
\end{center}

\bigskip
\normalsize

\textbf{Authors:} 
Joaquin Gonzalez-Nuevo$^{1,2}$ (gnuevo@uniovi.es); Laura Bonavera$^{1,2}$; Juan Alberto Cano$^{1,2}$; David Crespo$^{1,2}$; Rebeca Fernández-Fernández$^{1,2}$; Valentina Franco$^{1,2}$; Marcos M. Cueli$^{1,2}$; José Manuel Casas$^{1,2}$; Tony Mroczkowski$^3$; Caludia Ciccone$^4$; Evanthia Hatziminaoglou$^{5,6,7}$; Hugo Messias$^{8,9}$\\
$^1$Departamento de Fisica, Universidad de Oviedo, C. Federico Garcia Lorca 18, 33007 Oviedo, Spain.\\
$^2$Instituto Universitario de Ciencias y Tecnologías Espaciales de Asturias (ICTEA), C. Independencia 13, 33004 Oviedo, Spain.\\
$^3$Institute of Space Sciences (ICE, CSIC),Carrer de Can Magrans, s/n, 08193 Cerdanyola del Vallès, Barcelona, Spain.\\
$^4$Institute of Theoretical Astrophysics, University of Oslo, Postboks 1029, Blindern, 0315 Oslo\\
$^5$ESO, Karl-Schwarzschild-Str. 2, 85748 Garching bei München, Germany\\
$^6$Instituto de Astrof\'isica de Canarias (IAC), E-38200 La Laguna, Tenerife, Spain\\
$^7$Departamento de Astrof\'isica, Universidad de La Laguna, E-38206 La Laguna, Tenerife, Spain\\
$^8$European Southern Observatory, Alonso de C\'ordova 3107, Vitacura, Casilla 19001, Santiago de Chile, Chile\\
$^9$Joint ALMA Observatory, Alonso de C\'ordova 3107, Vitacura 763-0355, Santiago, Chile\\


\textbf{Science Keywords:} 
cosmology: large-scale structure of universe, cosmology: dark matter, cosmology: lenses, cosmology: observations

 \captionsetup{labelformat=empty}
\begin{figure}[h]
   \centering
   \captionsetup{width=.55\textwidth}
\includegraphics[width=.55\textwidth]{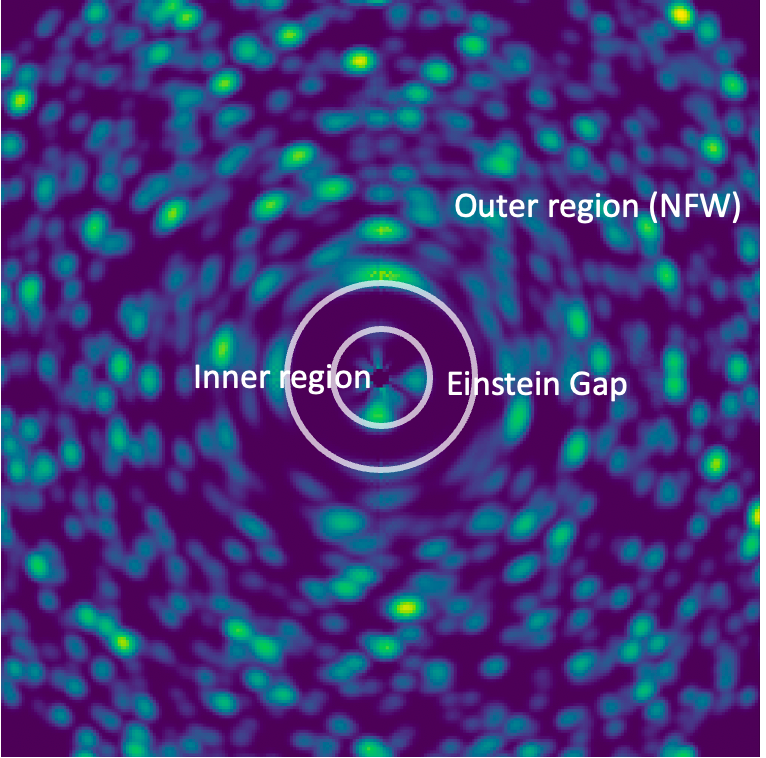}
   \caption{Simulated stacking map showing the strong lensing effect and, in particular, the gap inside the Einstein ring.}
\end{figure}
\vspace{-15mm}

\setcounter{figure}{0}
\captionsetup{labelformat=default}


\pagebreak
\section*{Abstract}

Magnification bias, the lensing-induced modification of background source number counts, provides a uniquely powerful probe of the mass density profiles of galaxies and clusters down to kpc scales. Unlike shear-based weak lensing, magnification bias does not rely on galaxy shapes and thus avoids dominant small-scale systematics. Existing studies, however, are limited by sky coverage, positional uncertainty, and insufficiently deep, confusion-limited submillimetre (submm) surveys. A next-generation wide-field, high-throughput submm facility like the proposed 50m-telescope AtLAST is required to unlock this technique’s full diagnostic power.

\section{Scientific context and motivation}
The internal structure of dark matter (DM) halos at $\sim 1$–$10$ kpc remains a major open question in astrophysics. Small-scale density profiles encode the interplay between baryonic processes, halo assembly, and the nature of DM, while at larger scales satellite galaxies influence the overall mass distribution (del Popolo \& Le Delliou, 2017). Observations also show that cluster substructures act as more efficient lenses than predicted (Meneghetti et al. 2020), suggesting DM subhalos may be more compact than standard models assume.
Weak-lensing shear has revolutionised our view of halos at $>100$ kpc, but it becomes severely limited at smaller radii due to shape noise and systematic uncertainties, making sub-10 kpc measurements unattainable even for next-generation facilities like \href{https://pole.uchicago.edu/public/Home.html}{\underline{SPT-3G}}, \href{https://simonsobservatory.org}{\underline{SO}}, and  \href{https://www.ccatobservatory.org}{\underline{FYST}}. Strong lensing probes these scales more effectively but requires precise constraints on multiple-image positions and redshifts, restricting its application to a small number of massive clusters (Umetsu, 2020; Fox et al., 2022).

Magnification bias offers a powerful complementary pathway. Instead of relying on galaxy shapes, it measures changes in background source counts via the angular cross-correlation between lenses and high-redshift submm galaxies (SMGs), whose steep number counts ($\beta>3$) maximise the magnification signal. Because SMGs and lenses are observed in separate bands, magnification bias is in principle limited only by positional accuracy, and stacking techniques allow studies of less massive halos (Crespo et al., 2024).
Recent work has shown that magnification bias can robustly reconstruct halo profiles for quasars, galaxies, and clusters. Stacking SMGs with improved astrometry (e.g. WISE, $\sigma\simeq0.3''$) now achieves kiloparsec-scale resolution, revealing a central excess, an outer power-law regime, and a deficit at $\sim10''$ ($10$–$60$ kpc) known as the “Einstein Gap.” An additional oscillatory feature at larger radii has been traced to the strong-lensing influence of massive satellites, which can bias Stellar-to-Halo Mass Ratio estimates for low-mass systems. Ray-tracing simulations confirm that the gap arises from strong-lensing–induced displacements, highlighting magnification bias as a sensitive probe of small-scale halo structure (Crespo et al., 2025).

Magnification-bias–based mass profile measurements remain limited primarily by the quality and coverage of current submm data. The small sky area available for high-redshift SMGs reduces statistical power, amplifies sampling variance, and weakens the signal at large angular scales, particularly in high-resolution analyses. As a result, current datasets limit the precision and cosmological reach of magnification-bias studies despite the technique’s demonstrated potential (Fernandez-Fernandez et al., 2025). A transformative leap in wide-field submm sensitivity and mapping speed is required to turn magnification bias into a precision small-scale lensing tool.

\vspace{-3mm}
\section{Science case}
A next-generation magnification bias programme directly addresses three high-impact questions in small-scale cosmology:

\textbf{The halo mass density profile at sub-10 kpc.} A central question in galaxy–halo physics is whether typical galaxy- and group-scale halos are intrinsically cuspy, following a NFW-like (Navarro et al., 1996) profile, or whether they instead host kiloparsec-scale cores shaped by baryonic feedback or alternative DM physics. In massive systems, energetic baryonic processes—stellar winds, supernovae, AGN outflows—can redistribute DM and produce shallower inner profiles, whereas smaller halos are expected to retain steeper, cuspier profiles (Faucher-Giguère et al., 2011; Somerville \& Davé, 2015). The detailed inner slope may also depend on the amount and distribution of baryonic material, including variations in the stellar initial mass function, which modulate the gravitational impact of the luminous component (Grillo, 2012). Directly probing these regimes remains challenging: conventional weak-lensing shear cannot reach such small scales, and strong lensing only provides constraints for a limited number of massive clusters (Zitrin et al., 2015). Magnification bias, however, uniquely offers access to sub-10 kpc radii across a broad range of lens types (Crespo et al., 2024). A large, homogeneous SMG survey with improved positional accuracy would enable systematic measurements of inner halo profiles at different masses and redshifts, opening a new observational window onto the interplay between baryons, DM, and possible departures from the standard CDM scenario.
\begin{wrapfigure}{r}{0.4\textwidth}
\includegraphics[width=.9\linewidth]{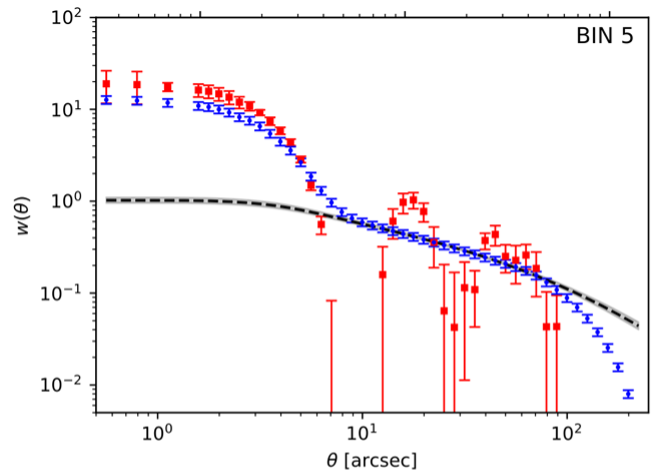}
   \caption{Cross-correlation functions measured using the magnification bias of SMGs (red dots) showing the lensing gap around $10''$ and the oscillatory signal at larger angular scales. The blue dots are the renormalized mass density profile measured from the galaxy satellites distribution. The best NFW fit to the outer region is represented by a dashed black line. See Crespo et al. (2025) for more details.}
\end{wrapfigure}
\textbf{    What causes persistent lensing anomalies in less massive galaxy clusters?} Current magnification bias measurements show an excess lensing signal and oscillatory features at larger separations (>100 kpc) in less massive clusters ($M_\star < 10^{10.8} M_\odot$, see Fig. 1 and Crespo et al. 2025). These less massive clusters cannot be studied with shear-based methods using currently available data. Both anomalies appear closely related and arise from a greater than expected strong lensing influence of massive satellite galaxies within foreground halos (Meneghetti et al., 2020, 2023). In low-mass systems, these satellites dominate the outer lensing signal, leading to inflated halo mass estimates when fitting standard profiles such as NFW. Moreover, their stochastic cumulative effect produces oscillatory patterns in the observed profiles, revealing that magnification bias primarily traces substructure rather than the much weaker smooth DM distribution. To derive a more robust interpretation of these anomalies in such low-mass halos, large, homogeneous surveys of SMGs are needed to overcome current statistical and systematic limitations.

\textbf{Determining the splashback radius for less massive galaxy clusters.} The splashback radius corresponds to the orbital apocenter of accreted matter, defining the physical boundary of DM halos. Characterising the splashback radius is paramount because it provides a powerful probe for constraining fundamental theories of structure formation and the nature of DM itself (Umetsu, 2020). Contigiani et al. (2019) and Umetsu \& Diemer (2017) were able to determine it through shear lensing observations but using similarly high-mass clusters. Magnification bias characteristics will allow us to determine such important quantity for a sample of less massive clusters or different kinds of lenses, where shear measurements are more difficult. To derive precise determinations of the splashback radius, large, homogeneous surveys of SMGs are needed.

Reaching the sub-10 kpc regime with magnification bias requires exceptionally precise positional information for background sources, as the achievable angular resolution is fundamentally limited by catalogue accuracy. Current analyses obtain few-kpc resolution by cross-matching SMGs with WISE, where the $\sim0.3''$ positional uncertainty sets the practical limit but introduce miss-matching issue when multiple counterparts are available. Pushing to smaller radii and avoiding cross-matching uncertainties will therefore require future catalogues with substantially improved intrinsic astrometric precision of $<0.5''$. The use of SMGs as background sources remains essential, since their steep number counts ($\beta \geq 3$) maximise the magnification response and make small-scale measurements feasible.
However, precision alone is not sufficient. Present magnification-bias studies are severely constrained by the limited sky coverage of existing submm surveys, which reduces statistical power and weakens constraints across all angular scales. While high resolution governs access to the innermost regions, wider and more homogeneous survey areas are crucial for stabilising the signal at large separations and for improving the overall robustness of halo-profile measurements. 

Current and planned facilities cannot meet the requirements for precision magnification bias measurements because none combine high sensitivity with wide-area coverage at submm wavelengths. \href{https://almascience.eso.org}{\underline{ALMA}}’s tiny field of view and slow survey speed make large-scale mapping impractical, while \href{https://www.eaobservatory.org/jcmt/instrumentation/continuum/scuba-2/}{\underline{SCUBA-2}}, \href{https://pole.uchicago.edu/public/Home.html}{\underline{SPT-3G}}, \href{https://simonsobservatory.org}{\underline{SO}}, and planned 6-m class telescopes (e.g. \href{https://www.ccatobservatory.org}{\underline{FYST}}) lack the resolution and sensitivity to increase background source density or reduce confusion noise.
Future facilities capable of combining high astrometric accuracy with broad sky coverage will therefore be key to unlocking the full potential of magnification bias, enabling precise halo studies from kiloparsec scales to the outskirts of galaxy systems.

\section{Technical requirements}
No existing or planned facility can deliver the capabilities required to transform magnification bias from a niche technique into a precision tool for mapping halo profiles from 1–200 kpc, comparing lens populations, and identifying anomalies that may signal new physics. Achieving these goals demands a telescope that combines wide-area coverage, high angular resolution, and exceptional sensitivity across the submm range (van Kampen et al., 2025). The proposed Atacama Large Aperture Submillimeter Telescope (\href{https://www.atlast.uio.no}{\underline{AtLAST}}; Mroczkowski et al., 2025) exemplifies these requirements. Its 50-meter aperture provides diffraction-limited resolution of about 1.5'' at 950 GHz, effectively eliminating confusion noise and enabling the detection of faint, dust-obscured galaxies (Mroczkowski et al., 2025). Its expected astrometric precision of $\lesssim 0.5''$ is critical for probing the innermost halo regions at kiloparsec scales. AtLAST’s instantaneous two-degree field of view, coupled with mapping speeds up to $10^3-10^5$ times faster than ALMA, makes surveys of thousands of square degrees feasible—essential for building statistically robust lens–source cross-correlation samples.

Next-generation instrumentation on AtLAST will include large-format cameras with over one million pixels, delivering unprecedented sensitivity and dramatically increasing the surface density of high-redshift submm galaxies. This will boost the signal-to-noise ratio of magnification-bias measurements by orders of magnitude. Its broad frequency coverage (30–950 GHz; Mroczkowski et al., 2025) enables accurate colour selection and photometric redshifts for background sources, while simultaneous spectroscopic and continuum observations will allow detailed characterization of lens and source populations (Gallardo et al., 2024). These combined capabilities represent a transformative leap beyond current facilities, enabling science that cannot be achieved by ALMA, \href{https://science.nasa.gov/mission/webb/}{\underline{JWST}}, \href{https://www.cosmos.esa.int/web/euclid}{\textit{\underline{Euclid}}} or \href{https://www.lsst.org}{\underline{LSST}}. Last but not least, AtLAST is pioneering sustainable astronomy: it is the first astronomical facility that, since its initial design phase, has been researching engineering solutions for a tailored, off-grid renewable energy system inclusive of a hybrid battery/hydrogen storage that can supply 100\% of the power needed for day and night-time telescope operations (Viole et al., 2023; Kiselev et al., 2024).

\bigskip
\noindent\textbf{References:} \small{Crespo et al., 2024, A\&A, 684, A109 $\bullet$ Contigiani, O., et al., 2019, MNRAS, 485, 408 $\bullet$ Crespo et al., 2024, arXiv:2509.02213 $\bullet$  del Popolo, A. \& Le Delliou, M., 2017, Galaxies, 5, 17 $\bullet$ Faucher-Giguère, A. et al., 2011, MNRAS, 417, 2982 $\bullet$ Fernandez-Fernandez, R., et al., 2025, arXiv:2510.23582  $\bullet$ Fox, C., et al., 2022, ApJ, 928, 87 $\bullet$ Gallardo, P.A., et al., 2024, Proc. of the SPIE, 13094, 1309428 $\bullet$ Grillo, C., 2012, ApJ Letters, 747, L15 $\bullet$ van Kampen, E., et al., 2025, Open Research Europe, 4, 122 $\bullet$ Kiselev, A., et al., 2024, Proc. of the SPIE, 13094, 130940E $\bullet$ Meneghetti, M., et al., 2020, Science, 369, 6509, 1347 $\bullet$ Meneghetti, M., et al., 2023, A\&A, 678, L2 $\bullet$ Mroczkowski, T., et al., 2025, A\&A, 694, A142 $\bullet$ Navarro, Frenk \& White1996, ApJ 462, 563 $\bullet$ Somerville, R. S. \&  Davé, R., 2015, ARA\&A, 53, 51 $\bullet$ Umetsu, K. \& Diemer, B., ApJ, 836, 231 $\bullet$ Umetsu, K., 2020, A\&A Rev., 28, 7 $\bullet$ Viole, I., et al., 2023, Energy, 282, 128570 $\bullet$ Zitrin, A., et al., 2015, ApJ, 801, 44}

\end{document}